\documentclass[11pt,twoside]{article}
\usepackage{newpasp}

\def\edcomment#1{\iffalse\marginpar{\raggedright\sl#1\/}\else\relax\fi}
\reversemarginpar

\begin{document}
\title{Deflection of jets induced by jet--cloud interactions}
\author{S Mendoza \& M S Longair}
\affil{Cavendish Laboratory, Madingley Road, Cambridge CB3 OHE, UK }

\begin{abstract}
  A non--relativistic and relativistic model in which astrophysical
jets are deflected on passing through an isothermal high density region
is analysed.  The criteria for the stability of jets due to the formation
of internal shocks are discussed.  \end{abstract}

\section{Introduction}
  The standard model for Fanaroff-Riley type II radio sources is described
as a pair of relativistic jets of electrons which is ejected in opposite
directions  from a very compact region in the nucleus of an elliptical
galaxy \cite{blandford90}.  The jets expand adiabatically through the
interstellar medium of the host galaxy and the surrounding intergalactic
medium.  Collimation of the jets is provided by the presence of a
cavity or ``cocoon'' surrounding the jet, which maintains the pressure
of the intergalactic medium in balance with that of the relativistic gas
within the jet.  In this model, the jets maintain a straight trajectory.
However, observations have shown that jets often bent.

  Typical examples of bending in radio sources appear in \emph{radio
trail} sources, \emph{mirror symmetric} sources and radio galaxies
which present \emph{inversion symmetry} \cite{begelman84}. Another way
of inducing deflections in radio jets is if the beam passes through
a region of interstellar or intergalactic gas in which there is a
significant pressure gradient.  This can occur if the jets pass through
the interstellar medium of a large galaxy or diffuse intergalactic
cloud.  Observations of this might be present in the radio galaxy
3C31 \cite{best97a}, but the quasar PKS 1318 + 113 \cite{lehnert99}
is an extremely good evidence of a deflection induced by a jet--cloud
interaction.

\section{Jet--cloud interactions}

  The interaction of an astrophysical jet with a cloud, in which the
characteristic size of the cloud is much greater than the jet radius,
has been studied in its initial stages by Raga \& Cant\'o (1995).
Their results show that initially the jet slowly begins to bore a passage
into the cloud.  Eventually a stationary situation is achieved in which
the jet  penetrates the cloud and escapes from it in a direction which
is different from the original jet trajectory.

  Once the steady state in this interaction is reached, the trajectory
is determined by maintaining pressure equilibrium with the surrounding
environment.  Since the expansion of the jet is adiabatic and a steady
state has been reached, Bernoulli's equation in its relativistic and
non--relativistic form \cite{daufm} can be used to describe the
path of the jet \cite{mendoza00b}.  

  With all these considerations, once the distribution of the gas inside
the cloud is known, it is possible to integrate Bernoulli's equation
numerically with the use of a Runge--Kutta method.  Analitic solutions are
possible for the case in which the jet penetrates the cloud at a position
\( \mathbf{r}_0 \) with a high supersonic motion \cite{mendoza00b}.

  Let us first discuss the case in which the cloud is an isothermal sphere
for which it's density, and hence its pressure, is inversely proportional
to the square of the radius vector.  Under these circumstances,
the problem is characterised by the gravitational constant \( G \),
a ``\textsf{characteristic length}'' \( r_0 \) and the values of the
velocity of the jet and the density at this point which are \( v_0 \) and
\( \rho_0 \) respectively.  Three independent dimensions (length, time and
mass) describe the whole hydrodynamical problem.  Since four independent
physical quantities (\( G,\, \rho_0,\, v_0,\,\textrm{and}\, r_0 \)) are
fundamental for the problem we are interested, the \mbox{Buckingham \(
\Pi \)--Theorem} \cite{sedov93} of dimensional analysis demands the
existence of only one dimensionless parameter \( \Lambda \equiv G \rho_0
r_0^2 / v_0^2  \).  This parameter is obtained naturally from the full
analytic solution \cite{mendoza00b}.  The other obvious dimensionless
number that parametrises the solutions is the Mach number \(M_0\)
evaluated at position \( r_0 \).

  The deflection of non--relativitic jets in isothermal clouds might
be important for interstellar molecular clouds and the jets associated
with Herbig--Haro objects.  If we adopt  a particle number density of \(
n_\mathrm{H} \! \sim \!  10^2 \, \textrm{cm}^{-3} \), and a temperature \(
T \! \sim 10 \,  \textrm{K}  \)  for a cloud with radius \( r_0 \! \sim
1 \, \textrm{pc} \) \cite{spitzer98,hartmann98}, then \( \Lambda  \sim
10^ {-2} / M_0^2 \). The same calculation can be made for the cases
of radio jets interacting with the gas inside a cluster of galaxies.
For this case, typical values are \( n_\mathrm{H} \sim 10^{-2} \,
\textrm{cm}^{-3} ,\ T \sim 10^7 \, \textrm{K}  \textrm{ and } r_0 \sim
100 \, \textrm{Kpc} \) (\cite{HEA2,galaxyformation}).  With these values,
the parameter  \( \Lambda \sim 10^{-2} / M_0^2 \), which is the same
value as the one obtained for Herbig--Haro objects.  The fact that jets
are formed in various environments such as giant molecular clouds and
the gaseous haloes of clusters of galaxies with the same values of the
dimensionless parameter \( \Lambda \) provides a clue as to why the jets
look the same in such widely different environments.

  From its definition, the parameter \( \Lambda \) can be rewritten as \(
\Lambda = ( 3 / 4 \pi ) \times  ( G { \mathsf{M} } / r ) ( 1 / v_0^2 )
\), where \( { \mathsf{M} } \) is the mass within a sphere of radius \(
r_0 \).  This quantity is roughly the ratio of the gravitational potential
energy from the cloud acting on a fluid element of the jet, to its kinetic
energy at the initial position \( r_0 \). The parameter \( \Lambda \)
is thus an indicator of how large the deflections due to gravity are
going to affect the trajectory of the jet. The bigger the number \(
\Lambda \), the more important the deflection caused by gravity will be.
In other words, when the parameter \( \Lambda \gg 1 \) the jet becomes
ballistic and bends towards the centre of the cloud.  When \( \Lambda
\ll 1 \) the deflections are dominated by the pressure gradients in the
cloud and the jets bend away from the centre of the cloud.

    Let us consider next the case of a galaxy dominated by a dark
matter halo for which the mass density is given by the relation
\cite{binney87}: \(  \rho_d = \rho_{d_\star} / \left( 1 + \left(
r / a \right)^2 \right) \), in which \( a \) is the core radius and
quantities with a star refer to the value at the centre of the galaxy.
The gas in the galaxy is in hydrostatic equilibrium with the dark matter
halo, so that \( { \mathbf{grad} }\: p \! = -\rho \, { \mathbf{grad}
}  \: \phi_d \).  In exactly the same form as it was done above, the
important parameters in the problem are the gravitational constant \(
G \), the characteristic length \( a \) and the sound speed \( c_\star
\), together with the gas density \( \rho_\star \) evaluated  at the
centre of the cloud.  Since three independent dimensions (length,
time and mass) describe the whole hydrodynamical problem, dimensional
analysis demands the existence of only one dimensionless parameter \(
\mathit{k} \equiv - 4 \pi G \rho_{\mathrm{d}_\star} a^2 / c_\star^2 \).
The unimportant proportionality factor of \( -4\pi \) is introduced here
since it appears naturally in the analytic solution \cite{mendoza00b}.
The Mach number \( M_0 \), evaluated at the point where the jet enters
the cloud is another dimensionless number which parametrises the solution
of this problem.  Using typical values \cite{binney87} for galaxies
then \( \rho_\star \! \sim \!  0.1 \, M_\odot \, \textrm{pc}^{-3} \),
\( \mathsf{a} \! \sim \! 1 \, \textrm{Kpc}  \) . Taking central values
for the gas in the galaxy as \( n_\star \! \sim \! 1 \textrm{cm}^{-3}
\) and \( T_\star \! \sim \!  10^5 \textrm{K} \) then \( \mathit{k}
\! \sim \! -10 \).

  The number \( \mathit{k} \) can be rewritten as \( \mathit{k} \! =
\! - \left( 4 / ( 4 - \pi ) \right)  ( G { \mathsf{M} } / \mathsf{a}
) ( 1 / c_\star^2 ) \), where \( { \mathsf{M} } \) is the mass of a
sphere with radius \( \mathsf{a} \).  This quantity is proportional
to the gravitational energy of the cloud evaluated at the core radius
divided by the sonic kinetic energy that a fluid element in the jet has.
In other words, in the same way as it was done above, the dimensionless
number \( \mathit{k} \) is an indicator as to how big deflections produced
by gravity are.

  Consider now the case in which relativistic effects are included in the
interaction  of a relativistic jet and a stratified high density region.
To simplify the problem, the self gravity of the cloud acting on the jet
is ignored.  Bernoulli's equation in its non--relativistic form can be
integrated numerically under this considerations and it is possible to
find out analytic solutions when the Mach number of the flow inside the
jet is much greater than unity.  This was done for the case in which the
cloud is an isothermal sphere and for the case in which the interaction
is between the jet and gas in pressure equilibrium with a dark matter
halo \cite{mendoza00b}.

\section{Discussion}

  When a jet bends it is in direct contact with its surroundings and
entrainment from the external gas might cause disruption to its structure
\cite{icke91}. However, if this situation is bypassed for example by
an efficient cooling, then there remains a high Mach number collimated
flow inside a curved jet. When supersonic flow bends, the characteristics
emanating from it intersect at a certain point in space \cite{daufm}.
Since the hydrodynamical values of the flow in a characteristic line
have constant values, the intersection causes the different values of
these quantities to be multivalued.  This situation can not occur in
nature and a shock wave is formed.

  The formation of internal shocks inside the jet gives rise to
subsonic flow inside it  and collimation may no longer be achieved.
If the characteristic lines produced by the flow inside the jet intersect
outside it, then a shock wave is not formed and the jet remains collimated
as it bends.  However, two important points have to be considered in
the discussion.  The first is that the Mach number decreases in a bend
as the flow moves.  The second is that the rate of change of the Mach
angle with respect to the angle the jet makes with its original straight
trajectory (the bending angle)  increases without bound as the velocity
of the flow tends to that of the local velocity of sound.  This was first
proved by Icke (1991) for the case in which no relativistic effects
were taken into account.  We have made a relativistic generalisation to
these two points \cite{mendoza00c}.

  The fact that two shocks might form inside a jet, one of them at
the end of the bending when the Mach number is near one, enables us to
find an upper limit to the bending angle \cite{icke91}.  For example, a
non--relativitic jet with a polytropic index \(\kappa \! = \! 5/3\) cannot
be deflected more than \(60.8^\circ\).  Under the same non--relativitic
conditions, but by assuming a polytropic index \(\kappa \! = \! 4/3\),
non--relativitic jets with a relativistic equation of state cannot be
deflected more than \(148.12^\circ\) \cite{icke91}.

  When a full relativistic analysis is introduced in this description,
this upper limit can not be greater than its non--relativitic counterpart.
This is because characteristic lines emanating from a relativistic flow
are closer to the streamlines as compared to their non--relativitic
counterparts \cite{konigl80}.  As a result, a relativistic jet with
a polytropic index \( \kappa \! = \! 4/3 \) can not bend more than
\(47.94^\circ\).  The precise conditions under which an internal shock
is produced for a given jet depend on the shape of the curve that the
jet makes as it bends and the radius of the jet.

  The most important consequence of the calculations described above is
the sensitivity  of the deflection angles to variations in velocity.
This sensitivity is due to the fact that the force applied to a given
fluid element in the jet (due to pressure and gravitational potential
gradients) is the same independent of the velocity of the flow in the
jet. However, as the velocity of the flow in the jet increases, there
is not enough time for this force to change the curvature of the jet
soon enough, giving rise to very straight jets.

\thebibliography

   \bibitem[(Begelman, Blandford, \& Rees 1984)]{begelman84}  Begelman M.C.,
        Blandford R.D., \&  Rees M.J. 1984, `Theory of extragalactic
        radio sources' Rev. Mod. Phys. 56, 255

   \bibitem[(Best, Longair, \& R\"ottgering 1997)]{best97a} Best P.N.,
        Longair M.S., R\"ottgering H.J.A. 1997, `A jet-cloud interaction
        in 3C 34 at redshift z=0.69' , \mnras 286, 785

   \bibitem[(Binney \& Tremaine 1987)]{binney87}
        Binney J., \&  Tremaine S. 1987, `Galactic Dynamics' (New
        Jersey: Princeton University Press)

   \bibitem[(Blandford 1990)]{blandford90} Blandford
           R.D. 1990,  in `Active Galactic Nuclei', ed. Courvoisier
           T. L. \& Mayor M.,  Saas-Fee Advanced Course 20 (Les
           Diablerets: Springer-Verlag), 161--275

   \bibitem[(Hartmann 1998)]{hartmann98} Hartmann L.
           1998, `Accretion Processes in Star Formation', Cambridge
           astrophysics series 32, (Cambridge, UK :  CUP)

   \bibitem[(Icke 1991)]{icke91} Icke V. 1991, `From Nucleus to
           Hotspot', in `Beams and Jets in Astrophysics', ed. Huges,
           P.A. (Cambridge, UK: CUP)

   \bibitem[(K\"onigl 1980)]{konigl80} K\"onigl A., 1980, `Relativistic
           gas dynamics in two dimensions', Physics of Fluids, 23, 1083

   \bibitem[(Landau \& Lifshitz 1987)]{daufm} Landau
           L.D., \& Lifshitz, E.M. 1987, `Fluid Mechanics' (London, UK:
           Pergamon)

   \bibitem[(Lehnert, et al.)]{lehnert99} Lehnert, M.D. \& van
           Breugel, W.J.M., Heckman T.M., \& Miley, G.K. 1999,
           `Hubble Space Telescope Imaging of the Host Galaxies of
           High-Redshift Radio-loud Quasars', \apjs, 124, 11

   \bibitem[Longair 1992]{HEA2} Longair, M.S. 1992, `High 
           Energy Astrophysics' (Cambridge, UK: CUP)

   \bibitem[Longair 1998]{galaxyformation} Longair, M.S. 1998, 
           `Galaxy Formation' (London, UK: Springer)

   \bibitem[(Mendoza 2000)]{mendoza00c} Mendoza, S. 2000, `Shocks and Jets 
           in Radio Galaxies and Quasars', PhD thesis, Cavendish
           Laboratory, University of Cambridge, UK

   \bibitem[(Mendoza, \& Longair  2000)]{mendoza00b} Mendoza, S., \&
           Longair, M.S.  2000, `Deflection of jets induced by jet-cloud
           \& jet-galaxy interactions', astro-ph/008015. Accepted for 
           publication in \mnras

   \bibitem[(Raga, \& Cant\'o 1995)]{raga95} Raga A.C., \&  Cant\'o J.
          1995, `The initial stages of an HH jet/cloud core collision',
           Rev. Mex. Astron. Astrofis., 31, 51

   \bibitem[(Sedov 1993)]{sedov93} Sedov, L 1993 `Similarity and
           Dimensional Methods in Mechanics', 10th edn (USA: CRC Press)

   \bibitem[(Spitzer 1998)]{spitzer98} Spitzer L.
           1998, `Physical Processes in the Interstelar Medium' (USA:
           Willey)

\endthebibliography

\end{document}